\documentclass[useAMS,usenatbib]{mn2e}

\usepackage{graphicx}
\usepackage{amsmath}
\usepackage{amssymb}
\def\aw{Alfv\'en}
\def\gp{$\Gamma^\prime$}

\title[Non-linear CR diffusion]{Non-linear diffusion of cosmic rays escaping from supernova remnants - I. The effect of neutrals}
\author[L. Nava et al.]
{L. Nava$^{1}$\thanks{E-mail: lara.nava@mail.huji.ac.il}, 
S. Gabici$^{2}$,
A. Marcowith$^3$,
G. Morlino$^4$,
and V.S. Ptuskin$^5$\\
$^{1}$Racah Institute of Physics, The Hebrew University of Jerusalem, 91904, Israel\\
$^{2}$APC, AstroParticule et Cosmologie, Universit\'e Paris Diderot, CNRS/IN2P3, CEA/Irfu, Observatoire de Paris, Sorbonne Paris Cit\'e,\\ 10, rue Alice Domon et L\'eonie Duquet, F-75205 Paris Cedex 13, France\\
$^{3}$Laboratoire Univers et particules de Montpellier, Universit\'e Montpellier/CNRS, F-34095 Montpellier, France\\
$^4$INFN -- Gran Sasso Science Institute, viale F. Crispi 7, I-67100 L'Aquila, Italy\\
$^5$Pushkov Institute of Terrestrial Magnetism, Ionosphere and Radio Wave Propagation of the Russian Academy of Sciences, Troitsk,\\ Moscow, 142190, Russia}
\begin{document}
\voffset -1truecm 
\date{}

\pagerange{\pageref{firstpage}--\pageref{lastpage}} \pubyear{}

\maketitle

\label{firstpage}

\begin{abstract}
Supernova remnants are believed to be the main sources of galactic Cosmic Rays (CR). Within this framework, particles are accelerated at supernova remnant shocks and then released in the interstellar medium. The mechanism through which CRs are released and the way in which they propagate still remain open issues. 
The main difficulty is the high non-linearity of the problem: CRs themselves excite the magnetic turbulence that confines them close to their sources.
We solve numerically the coupled differential equations describing the evolution in space and time of the escaping particles and of the waves generated through the CR streaming instability. The warm ionized and warm neutral phases of the interstellar medium are considered. These phases occupy the largest fraction of the disc volume, where most supernovae explode, and are characterised by the significant presence of neutral particles. The friction between those neutrals and ions results in a very effective wave damping mechanism. It is found that streaming instability affects the propagation of CRs even in the presence of ion-neutral friction. The diffusion coefficient can be suppressed by more than a factor of $\sim 2$ over a region of few tens of pc around the remnant. The suppression increases for smaller distances. The propagation of $\approx 10$ GeV particles is affected for several tens of kiloyears after escape, while $\approx 1$ TeV particles are affected for few kiloyears. This might have a great impact on the interpretation of gamma-ray observations of molecular clouds located in the vicinity of supernova remnants.
\end{abstract}


\section{Introduction}
\label{sect:}

Cosmic Rays (CRs), discovered more than one century ago, are known to propagate in our Galaxy by successive random walks with mean free paths which can exceed a parsec at energies above a few GeV. The origin of CRs is still a subject of debate. Though supernova remnants (SNRs) are the favourite sources \citep{drury2001}, a contribution from other types of sources like, for instance, superbubbles \citep{bykov2001, parizot2004} can not be excluded at this stage. In SNRs, diffusive shock acceleration appears to be an efficient mechanism to accelerate particles up to PeV energies or possibly beyond (\citealt{bell2013} and references therein). Unfortunately, a fully self-consistent theory describing both acceleration of particles at SNR shocks and CR propagation in the interstellar medium (ISM) is still missing. In particular, several important issues still remain opened: how do CRs escape from their sources? What is the impact of the properties of the ISM over CR propagation? This latter question is addressed in this article. 

Modelling the propagation of CRs just after their escape allows us to predict how their intensity varies as a function of time and space in the immediate vicinity of their accelerators.
The intensity of CRs close to an accelerator is expected to be larger than that of the CR background for a certain time-scale, which is determined by the diffusion coefficient of CRs in the local environment of the accelerator. Such scenario can be constrained observationally if CRs, during their propagation, interact with a dense medium (e.g. a molecular cloud) and generate gamma-rays in inelastic proton-proton interactions \citep{gabici07, gabici09, sabrina, ohira}. The recent detection of gamma-ray emission around some SNRs \citep[e.g.][]{W28hess,W28fermi,W44fermi}, and the advent of future facilities such as the Cherenkov Telescope Array (https://www.cta-observatory.org/) motivates the development of accurate models capable to describe the CR intensity nearby SNRs at different ages and propagating in different type of ISM.

The difficulty of the escape process resides in its non-linearity, since CRs themselves are believed to generate the magnetic turbulence needed to confine them. The role of CR self-confinement in and around CR sources is a long-standing issue already addressed in several articles \citep{skilling1975, bell78, schwartz78, ptuskinzirakashvili2005, plesser, lazarianW28, malkov}. The process, in principle 3 dimensional, can be treated using the so-called flux tube approximation where escaping CRs start to stream freely along the magnetic field lines of a flux tube containing the SNR shock \citep{plesser}. While streaming at velocities faster than the local Alfv\'en velocity, CR trigger magnetohydrodynamic waves \citep{skilling1975}. Among them, the slab waves which propagate along the magnetic field lines grow the fastest. The problem can be considered as one dimensional because magnetic field lines are only weakly perturbed up to a scale of the order of the coherence length of the background interstellar turbulence \citep{nava13}. Beyond this distance, a 3D transport process starts to control the magnetic field line transport. The turbulent coherence length is believed to be in the range $\approx$ 50-100 parsecs (see the recent results in \citealt{beck16} and references therein).
If 1D transport along the magnetic field lines prevails, then the CR transport and the wave growth (and possible damping) can be described by a system of two coupled non-linear equations. Waves and particles interact resonantly, i.e. the slab mode wavelength $k$ and the particle Larmor radius $r_{\rm L}$ verify $k r_{\rm L} \sim 1$ (see \S \ref{sec:problem}). In this model, the intensity of the perturbations at a given $k$ fixes the efficiency of the particle scattering and hence the normalisation of the spatial diffusion coefficient at the resonant CR energy. In order to derive analytical solutions of this non-linear system, \cite{plesser} and \cite{malkov} have considered self-similar solutions. They predict a space- and time-dependent CR distribution which can eventually be probed by gamma-ray experiments if a molecular cloud is standing at sufficiently small distances from the source. One of the major caveat of self-similar solutions is that if CRs produce strong wave generation they produce a strong particle confinement that should lead to unrealistically high gamma-ray fluxes. This aspect motivated the numerical approach exposed in this article. 

The aim of our work is many folds. At first, we proceed with a parametric survey of both SNR and local ISM properties and investigate under which particular conditions the growth of waves is effective. Secondly, we consider the role of wave damping depending on the type of ISM in which CRs propagate. In particular, we investigate the 
damping induced by the presence of neutral particles. For this reason, we limit our investigation to two phases of the ISM: the Warm Neutral Medium (WNM) and to the Warm Ionized Medium (WIM). \\

The layout of this article is as follows: Section \ref{sec:problem} introduces the mathematical and physical formalisms adopted in the flux tube approximation. In section \ref{sec:Num}, we present the numerical solutions of the CR escape problem including the effect of a linear damping term. In this section, we detail the impact of the initial CR pressure, damping rate and background turbulence level on the solutions. In section \ref{sec:ISM}, we specify the dominant damping process depending on the ISM phase and we propose an estimation of the typical escaping time-scale of CRs from their parent source. In section \ref{sec:Tim}, we discuss the astrophysical consequences of our model. We conclude in section \ref{sec:Con}.

\section{The problem}
\label{sec:problem}

The aim of this paper is to study the non-linear diffusion of a population of CRs after their escape from the acceleration site (for example, an SNR shock). As a zero--order approximation, the problem can be described by an impulsive injection of a given amount of CRs at a given location in the Galaxy. For simplicity, we consider here only relativistic CRs, with particle energies above $\approx 1$~GeV and thus velocities roughly equal to the speed of light $c$.

The transport of CRs is assumed to be regulated by the resonant scattering off Alfv\'en waves, i.e. a CR of energy $E$ resonates with waves of wave number $k = 1/r_{\rm L}$ \citep[see e.g.][]{wentzel}, where the Larmor radius $r_{\rm L}$ of a particle with Lorentz factor $\gamma$, mass $m$, and charge $Ze$ gyrating with velocity $v$ in a magnetic field of strength $B$ is $r_{\rm L}=\frac{\gamma m v c}{ZeB}$. For relativistic protons, $r_{\rm L}\simeq E/(eB)$. The (normalized) energy density $I(k)$ of Alfv\'en waves is defined as
\begin{equation}
\frac{\delta B^2}{8 \pi} = \frac{B_0^2}{8 \pi} \int I(k)\, {\rm d}\ln k ~ ,
\end{equation}
where $B_0$ is the ambient magnetic field and $\delta B$ the amplitude of the magnetic field fluctuations.

According to quasi--linear theory, CRs diffuse along the magnetic field lines with a diffusion coefficient equal to \citep{bell78}:
\begin{equation}
\label{eq:diff}
D(E) = \frac{4 ~ c ~ r_{\rm L}(E)}{3 \pi~ I(k)} = \frac{D_{\rm B}(E)}{I(k)} ~ ,
\end{equation}
which can be expressed as the ratio between the Bohm diffusion coefficient $D_{\rm B}(E)$ and the energy density of resonant waves $I(k)$. Formally, quasi--linear theory is valid for $\delta B/B_0 \ll 1$, and for a well-developed (i.e. broad) turbulent spectrum, which is the case considered here. In this limit, the diffusion perpendicular to the field lines can be neglected, being suppressed by a factor of $(\delta B_k/B_0)^4 \equiv I(k)^2$ with respect to the one parallel to $B_0$ \citep[see e.g.][]{drury1983,fabien}. This implies that under the condition that $\delta B/B_0$ remains small, the problem is one--dimensional. We verified a posteriori that this assumption is always satisfied in our calculations, and the choice to work in the limit of quasi--linear theory is hence well-justified. 

Finally, the additional assumption is made that the main source of Alfv\'enic turbulence is the streaming of CRs. Let us say that CRs stream along the direction of $B_0$, which is aligned along the coordinate $z$. Then the growth rate of Alfv\'en waves $\Gamma_{\rm CR}$ is proportional to the product between the \aw\ speed and the gradient of the pressure of resonant CRs, and can be defined as \citep[see e.g.][]{skilling1975,drury1983}:
\begin{equation}
\label{eq:growth}
2\Gamma_{\rm CR} I = - V_{\rm A} \frac{\partial P_{\rm CR}}{\partial z} ~ .
\end{equation}
For dimensional consistency, the pressure of the resonant CRs with energy $E$ ($P_{\rm CR}$ in Equation~\ref{eq:growth}) has been normalized to the magnetic energy density $B_0^2/8\pi$.
The right-hand side of the equation above can be interpreted as the rate of work done by the CRs in scattering off the waves, which is equal to the energy generation rate of waves (left-hand side of the equation). Thus, equation~\ref{eq:growth} implies a maximal efficiency for the growth of waves \citep[e.g.][]{drury1983,malkov} and its sign indicates that only waves travelling in the direction of the streaming are excited. 

It is now possible to write the equations that govern the transport of CRs and waves along the magnetic field lines. 
On scales smaller than the magnetic field coherence length, a flux tube characterized by a constant magnetic field strength $B_0$ and directed along the $z$--axis is considered.
The equations that describe the evolution of CRs and waves along the flux tube are coupled because the diffusion coefficient of CRs of energy $E$ (equation~\ref{eq:diff}) depends on the energy density of resonant waves $I(k)$, while the growth rate of waves (equation~\ref{eq:growth}) depends on the gradient of the pressure of resonant CRs $\partial P_{\rm CR}/\partial z$. The two equations then read:
\begin{equation}
\frac{\partial P_{\rm CR}}{\partial t} + V_{\rm A} \frac{\partial P_{\rm CR}}{\partial z} = \frac{\partial}{\partial z} \left( \frac{D_{\rm B}}{I} \frac{\partial P_{\rm CR}}{\partial z} \right)\,,
\label{eq:CRs}
\end{equation}
\begin{equation}
\frac{\partial I}{\partial t} + V_{\rm A} \frac{\partial I}{\partial z} = - V_{\rm A} \frac{\partial P_{\rm CR}}{\partial z} - 2\Gamma_{\rm d} I + Q\,,
\label{eq:waves}
\end{equation}
where the left hand side of both expressions is the time derivative computed along the characteristic of excited waves:
\begin{equation}
\frac{\rm d}{{\rm} d t} = \frac{\partial}{\partial t} + V_{\rm A} \frac{\partial}{\partial z} ~ .
\end{equation}
The advective terms $V_{\rm A} \partial P_{\rm CR}/\partial z$ and $V_{\rm A} \partial I/\partial z$ are neglected in the following since they play little role in the situation under examination. This assumption can be easily checked a posteriori.
The last two terms in equation~\ref{eq:waves} account for possible mechanisms of wave damping, operating at a rate $\Gamma_{\rm d}$, and for the injection $Q$ of turbulence from an external source (i.e. other than CR streaming). The term representing the external source of turbulence can be set to $Q =2 \Gamma_{\rm d} I_0$, so that when streaming instability is not relevant, the level of the background turbulence is at a constant level $I = I_0$. In general, the damping coefficient could be a function of $I$ \citep[see e.g.][]{plesser}, but for reasons that will be clear in the following, here we limit our analysis to linear damping terms, i.e. $\Gamma_{\rm d}$ is a constant in both space and time.

As pointed out by \cite{malkov}, the approach described above decouples the process of acceleration of particles (which operates, for example, at an SNR shock) from the particle escape from the acceleration site. In other words, equations~\ref{eq:CRs} and \ref{eq:waves} apply in the transition region between the acceleration site (where a high level of magnetic turbulence, typically at the Bohm level, may be expected) and the average conditions of the ISM, where the level of turbulence is much smaller. Though such a separation might seem artificial, the problem defined above can still be useful to describe the escape of particles from a {\it dead} accelerator, in which the acceleration mechanism does not operate anymore, or operates at a much reduced efficiency \citep{malkov}. This situation would probably apply to the case of old SNRs.

On the other hand, \cite{plesser} suggested that equations~\ref{eq:CRs} and \ref{eq:waves} could be also used to describe an intermediate phase of CR propagation in which the CRs have left the source but are not yet completely mixed with the CR background. For the case of an SNR shock, the equations above could thus be applied to those CRs characterized by a diffusion length large enough to decouple them from the shock region. Typically, this happens during the Sedov phase to the highest energy particles confined at the SNR shock when the diffusion length $D_{\rm B}/u_{\rm s}$ gradually increases with time  up to a value larger than some fraction $\chi$ of the SNR shock radius $R_{\rm s}$, where $u_{\rm s}$ is the shock speed and $\chi \approx 0.05,...,0.1$ \citep[e.g.][]{ptuskinzirakashvili2005,gabiciescape}.
In both the scenarios described above, CRs are decoupled from the SNR shock and we will refer to them as a {\it CR cloud}.

The initial conditions for equations~\ref{eq:CRs} and \ref{eq:waves} can be set as follows
\begin{eqnarray}
\label{eq:init_condP}
P_{\rm CR} &=& P_{\rm CR}^0 ~~~~~ z < a \\
	&=&  0 ~~~~~~~~~ z > a ~ ,
\end{eqnarray}
where $a$ represents the spatial scale of the region filled by CRs at the time of their escape from the source, and $I = I_0$ everywhere.
In fact, a larger value of $I \gg I_0$ could be chosen as an initial condition for $z < a$ (to mimic Bohm diffusion inside the accelerator). However, we found the exact initial value of $I$ inside the source to have very little effect on to the solution of the problem, and thus we kept a spatially uniform $I_0$.

Following \cite{malkov}, we introduce the quantity $\Pi$, defined as:
\begin{equation}
\Pi ~=~ \frac{V_{\rm A}}{D_{\rm B}} ~ \Phi_{\rm CR}\,,
\end{equation}
where:
\begin{equation}
\Phi_{\rm CR} \equiv \int_0^{\infty} P_{\rm CR}\,{\rm d}z = a\,P_{\rm CR}^0 ~ .
\end{equation}
It follows that $\Pi$ is a conserved quantity that can be considered as a control parameter.
To understand its meaning, consider the initial setup of the problem, in which CRs are localized in a region of size $a$. The CR pressure within $a$ is $P_{\rm CR}^0$, and then $\Phi_{\rm CR} = a\,P_{\rm CR}^0$. The initial diffusion coefficient outside the region of size $a$ is equal to $D_0=D_{\rm B}/I_0$.
In such setup three relevant timescales exist: i) the growth time of waves $\tau_g \approx I_0 /(V_{\rm A} \partial P_{\rm CR} /\partial z) \approx a I_0/V_{\rm A} P_{\rm CR}^0$, ii) the time it takes the CR cloud to spread significantly due to diffusion $\tau_{diff} \approx a^2 / D \approx a^2 I_0/D_{\rm B}$, and iii) the damping time $\tau_{damp} = 1/2\Gamma_d$. 
In order to have a significant growth of waves due to CR streaming, the time-scale for growth must be shorter than the two other time-scales:
$\tau_g<\min(\tau_{diff},\tau_{damp})$. In terms of the parameter $\Pi$, this conditions reads: $\Pi>\max(1,\tau_{diff}/\tau_{damp})$.
It is evident then that the parameter $\Pi$ regulates the effectiveness of CR--induced growth of waves. In the absence of a damping term for waves, $\Pi>1$ is a sufficient condition for streaming instability to be relevant, while in the presence of efficient wave damping, a more stringent condition on $\Pi$ applies. 

For $\Pi \lesssim \max(1,\tau_{diff}/\tau_{damp})$ CRs play no role in the generation of \aw\ waves, and equation~\ref{eq:CRs} can be solved analytically:
\begin{equation}
\label{eq:test}
P_{\rm CR} = \sqrt{\frac{I_0}{\pi D_{\rm B} t}} ~ \Phi_{\rm CR} ~ e^{-\frac{I_0 z^2}{4 D_{\rm B} t}}\,.
\end{equation}
Equation~\ref{eq:test} is referred to as {\it test--particle} (TP) solution.
When wave growth cannot be neglected, the solution deviates from the TP one.
It is worth noticing that, independently on the value of $\Pi$, equation~\ref{eq:test} is also the formal asymptotic solution of the non-linear equations~\ref{eq:CRs} and \ref{eq:waves} at late times, when CRs are spread over a large region along the magnetic flux tube, and consequently their ability to amplify Alfv\'en waves is strongly reduced (in other words, $\partial P_{\rm CR}/\partial z$ is small). 

For completeness, we mention that in the extreme scenario (not considered in this paper) where $\Pi\gg\max(1,\tau_{diff}/\tau_{damp})$ waves grow so quickly that the diffusive term in equation~\ref{eq:CRs} becomes negligible when compared to advection term $V_{\rm A} \partial P_{\rm CR}/\partial z$. In this case the advection term can no longer be neglected. This describes a situation in which CRs are "locked" to waves and move with them at a velocity equal to $V_{\rm A}$ \citep{skilling1971}. An identical result was found by \cite{cesarsky1975} in a study of the escape of $\approx$~MeV CRs from sources, and also suggested by \cite{hartquist}. 

To conclude this section, let us estimate the numerical values of $\Pi$, the main parameter that regulates the problem under examination. The value of $\Pi$ for the case of an SNR of radius $R_{\rm s} = 10~ R_{1}$~pc which releases $10^{50} W_{\rm CR,50}$~erg of CRs with a differential energy spectrum $\propto E^{-2.2}$ is:
\begin{equation}
\label{eq:pinum}
\Pi \approx 3 \times 10^4 ~ W_{\rm CR,50} ~ R_{1}^{-2} ~ n_{i,-1}^{-1/2} ~ E_1^{-1.2}
\end{equation}
where $n_{i,-1}$ is the density of the ionised gas in the ISM in units of 0.1 cm$^{-3}$ and $E_1$ the CR particle energy in units of 10 GeV.
This demonstrates that extremely large values of $\Pi$ (and thus of the initial growth rate) are very easily obtained, and explains why very small diffusion coefficients are obtained if the growth of waves is not balanced by a quite effective damping mechanism \citep[see e.g.][]{malkov}. 
On the other hand, in the presence of significant damping (i.e. when $\tau_{diff}/\tau_{damp} > 1$), the condition to be satisfied to have a growth of Alfv\'en waves is more stringent and reads $\Pi > \tau_{diff}/\tau_{damp}$. 
While in absence of significant damping large values of $\Pi$ lead to significant growth of waves and it is then easy to obtain values of $I\gg1$ (inconsistent with the initial assumption of quasi-linear regime), we will show that, in presence of damping, the derived values of $I$ always satisfy the condition $I\ll1$.

\section{Numerical solution: normalized equations}\label{sec:Num}

As shown above, the solution of Equations~\ref{eq:CRs} and~\ref{eq:waves} depends on the value of many physical parameters, including the CR energy and spectrum, the value of the interstellar magnetic field, the density of the ISM, the initial value of the background turbulence $I_0$, and the size of the region $a$ over which CRs are released. 
However, we show here that the main physical quantities in Equations~\ref{eq:CRs} and~\ref{eq:waves} can be conveniently re-normalised, so that the solution depends only on three variables. 
This study allows to understand how different initial conditions affect the final results, how the solution evolves with time and how the particle distribution in space is modified, as compared to the test-particle solution, in the cases where the growth rate is relevant.

First, we perform the following change of coordinates \citep{malkov}:
\begin{equation}
s\equiv\frac{z}{a}  \qquad\qquad  
\tau\equiv \left(\frac{V_{\rm A}}{a}\right) t~, 
\end{equation}
where the space and time coordinates are normalised to the values $a = (2/3) R_{\rm s} \approx 7 ~R_1$~pc\footnote{In our one-dimensional model the region initially filled with CRs has a volume equal to the transverse section of the flux tube enclosed by the SNR, $\pi R_{\rm s}^2$, times the thickness of the CR cloud along the flux tube, $2 a$. The effective size for $a$ is then obtained by equating this volume to $(4 \pi/3) R_{\rm s}^3$.} and $a/V_{\rm A} \approx 10^5 ~R_1 n_{i,-1}^{1/2} B_1^{-1}$~yr, where $B_1$ is the interstellar magnetic field strength in units of 10 $\mu$G.
Since the flux tube can be considered as a one-dimensional structure of length $L = s_{max} a$ equal to (at most) few hundred parsecs \citep[see e.g.][]{plesser}, the solutions presented in the following will be accurate up to a normalised spatial coordinate $s = s_{max}$ of the order of few tens to few hundreds.
Then, after performing the following normalisations:
\begin{eqnarray}
\mathcal{P}_{\rm CR} &\equiv& \left( \frac{V_{\rm A} a}{D_{\rm B}} \right) P_{\rm CR}~, \\
W &\equiv& \left(\frac{V_{\rm A} a}{D_{\rm B}} \right) I ~, \\
\Gamma^\prime &\equiv& \left( \frac{a}{V_{\rm A}} \right) \Gamma_{\rm d}~,
\end{eqnarray}
Equations~\ref{eq:CRs} and \ref{eq:waves} become:
\begin{eqnarray}
\frac{\partial \mathcal{P}_{\rm CR}}{\partial \tau} &=& \frac{\partial}{\partial s} \left( \frac{1}{W} \frac{\partial \mathcal{P}_{\rm CR}}{\partial s} \right) \label{eq:norm_p}~,\\
\frac{\partial W}{\partial \tau} &=& - \frac{\partial \mathcal{P}_{\rm CR}}{\partial s} - 2\Gamma^\prime \left( W - W_0 \right) \label{eq:norm_w}~.
\end{eqnarray}
Their solution (i.e., $\mathcal{P}_{\rm CR}$ and $W$ as a function of the variables $\tau$ and $s$) depends only on three parameters: the initial values $\mathcal{P}_{\rm CR}^0$, $W_0$, and $\Gamma^\prime_0$. Note that: $\mathcal{P}_{\rm CR}^0=\Pi$, $W_0=V_{\rm A}a/D_0$, and $\Gamma^\prime=\Gamma^\prime_0$ (since we consider here only a linear damping term which is constant in both time and space).
In this notation, the condition for effective growth of waves is $\Pi > \max(1,\Gamma^\prime W_0)$ and, similarly, the condition to have simultaneously growth of waves and still significant damping is $1 < W_0 \Gamma^{\prime}_{\rm d} \ll \Pi$.

In order to identify the relevant range of values for the parameters we notice that the (normalized) initial level of Alfv\'enic turbulence $W_0$ is generally expected to be much smaller than unity, otherwise the confinement time of CR close to the source would be unrealistically large. In fact $W_0$ can be written as the ratio of the characteristic diffusive time scale and the Alfv\'enic crossing time, which is clearly much smaller than unity: $W_0 = V_{\rm A} a/D_0 = (a^2/D_0) (a/V_{\rm A})^{-1} \ll 1$.
On the other hand, by recalling that $D_0 \sim \lambda c$, where $\lambda$ is the particle mean free path, we can write: $W_0 \sim (V_{\rm A}/c) (a/\lambda) \sim s_{max}^{-1} (V_{\rm A}/c) (L/\lambda)$. The condition for having a diffusive behaviour of CRs over a distance $L$ is $L/\lambda \gg 1$, which translates in a lower limit for $W_0$ which, for typical parameters, equals $\approx 10^{-6} - 10^{-5}$. 
We note that $\lambda < L$ is expected if the particles are able to generate slab modes, i.e. in the regime of efficient wave production.
Finally, the range of relevant values for $\Gamma_{\rm d}^{\prime}$ can be derived from the condition $1 < W_0 \Gamma^{\prime}_{\rm d} \ll \Pi$ and from the fact that $\Pi$ generally assumes very large values (see Equation~\ref{eq:pinum}).

\begin{figure*}
\hskip -0.55 truecm
\includegraphics[scale=0.95]{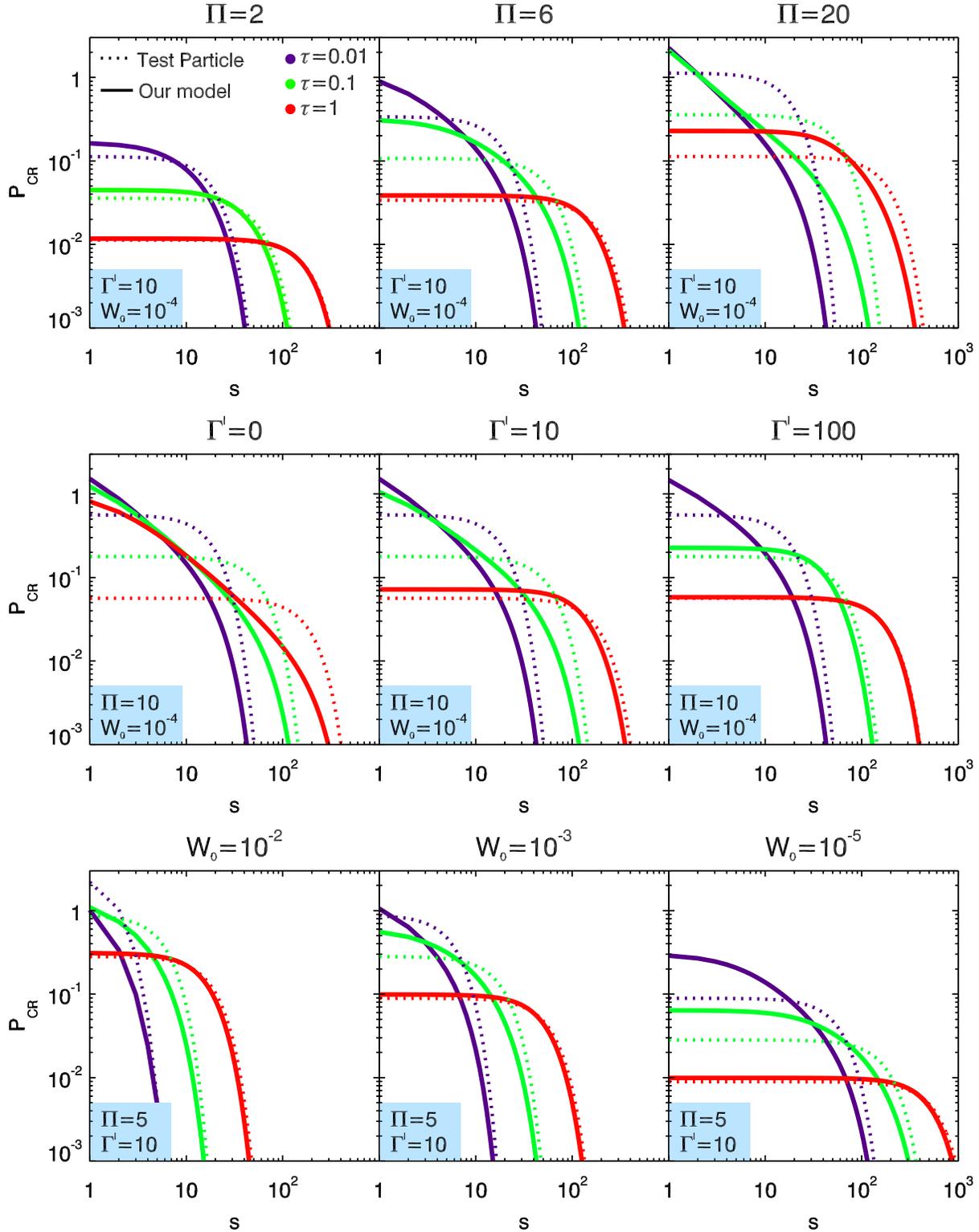}
\caption{Normalized CR pressure as a function of the variable $s$, which represents the distance from the source $z$ normalised to the size of the source $a$. In each row the results are shown by varying only one of the three key parameters (whose value is reported above each panel) and keeping the other two fixed to the values reported in the shaded area. The solid curves show the results from the numerical computation, while dotted curves represent the test particle solution, where the role of streaming instabilities in enhancing the turbulence is neglected. Different colours refer to different normalised times $\tau=10^{-2}, 10^{-1},1$.}
\label{fig:main}
\end{figure*}

\subsection{Results}
\label{sect:results}

We solved numerically Equations~\ref{eq:norm_p} and \ref{eq:norm_w} and show the results for $\mathcal{P}_{\rm CR}(s,\tau)$ as a function of the normalised coordinate $s$ (solid lines in Fig.~\ref{fig:main})
at three different normalised times $\tau=10^{-2},10^{-1},1$ (corresponding to different colours).
The test particle solution for the same values of $\Pi$ and $W_0$ is also shown for comparison (dotted lines).
Each row in the figure corresponds to results obtained by varying only one parameter at the time. 

The upper row shows the impact of the parameter $\Pi$ for fixed values of $\Gamma^\prime=10$ and $W_0=10^{-4}$. Since for $\Pi<1$ the numerical solution coincides with the TP solution, we show our results for values of $\Pi>1$. When $\Pi=2$, the solution starts to deviate from the test particle case: due to the growth of wave by streaming instability, the diffusion is slower, the CR pressure in the vicinity of the accelerator is larger, and the diffusion length is smaller. The effect is more and more evident for larger values of $\Pi$ (upper row, middle and right-hand panels).

The middle row shows the effect of changing the wave damping coefficient $\Gamma^{\prime}$, for fixed $\Pi=10$ and $W_0=10^{-4}$. Since $\Pi\gg1$, when wave damping is negligible (\gp=0, left-hand panel) the diffusion is suppressed as compared to the test particle case. Larger values of \gp\ limit the growth of the Alfv\'enic turbulence and the solution approaches the test particle solution. Note that \gp=10 (middle row, central panel) corresponds to a damping mechanism that becomes important at $\tau>0.1$. For this reason the solution at $\tau=0.01$ and $\tau=0.1$ are unaffected by the increased damping rate, when compared to the case \gp = 0, and only the solution at late time $\tau=1$ is modified, and approaches the test particle solution. The right-hand panel shows the case of an even faster damping mechanism, \gp=100, whose effect starts to be relevant at times corresponding to $\tau>0.01$, explaining why the solution at intermediate and late times approaches the test particle solution, while the solution at early time is unaffected by the increased damping rate.

The bottom row shows calculations performed at $\Pi=5$ and \gp=10, with $W_0$ varying from $10^{-2}$ to $10^{-5}$.
A large $W_0$ (left-hand panel) corresponds to slow diffusion and small diffusion length, both in the test particle case and in the numerical solution. The two solutions differ at early and intermediate times because the growth rate is faster than the diffusion rate, and the diffusion is suppressed. At later times, instead, the growth rate decreases and the numerical solution does not appreciably differ from the test particle case. A change in the initial value of the background turbulence has the same effect on both the analytical and numerical solution: the diffusion is faster for decreasing values of $W_0$ (see bottom row, from left to right), and the CRs can reach larger distances. The deviation from a test particle solution, however, remains unchanged, since both the initial growth rate and the diffusion rate scale as $W_0$.

\section{Applications to different phases of the interstellar medium}\label{sec:ISM}

We now apply the results obtained in the previous section to specific conditions of the ISM.
In particular, we focus on two different phases of the ISM: the warm neutral medium (WNM) and the warm ionised medium (WIM). 
The reason is that these two phases occupy the largest fraction of the volume of the disc, where most of the supernovae are expected to explode.
The cases of other interstellar phases will deserve a future work. Especially the case of the hot ionised medium is more complex as the dominant damping process might be
non-linear (see \citealt{plesser}), i.e. the damping coefficient might be a function of $I$.   
Our choice for the reference values of the physical quantities defining the properties of the WIM and WNM is reported in Table~\ref{tab:phases}.
All the values have been chosen within the ranges quoted by \cite{jean} (see their table 1).

\begin{table}
	\centering
	\caption{Reference physical quantities for the WIM and the WNM. $T$ is the gas temperature, $B_0$ the magnetic field strength, $n_{tot}$ the total (ion plus neutral) density and $f_{ion}$ the ion fraction.}
	\label{tab:phases}
	\begin{tabular}{c|cc} 
		\hline
		 & WIM & WNM \\
		\hline
		$T$ (K) & $8 \times 10^3$ & $8 \times 10^3$ \\
		$B_0$ ($\mu$G) & 5 & 5 \\
		$n_{tot}$ (cm$^{-3}$) & 0.35 & 0.35 \\
		$f_{ion}$ & 0.9 & 0.02 \\
		\hline
	\end{tabular}
\end{table}

\subsection{Damping of Alfv\'en waves}

Damping of the Alfv\'en waves can reduce the turbulence induced by streaming instability and hence reduce the self--confinement of CRs around their sources.
\aw\ waves are subject to various damping processes in the ISM. Which process is the dominant one depends dramatically on the properties of the medium.
In the two phases of the ISM considered here, the WIM and the WNM, the two most relevant damping mechanisms are the ion-neutral friction \citep{kulsrud71} and the damping by background MHD turbulence suggested by \citet{fg}, operating at a rate $\Gamma_{\rm d}^{IN}$ and $\Gamma_{\rm d}^{FG}$, respectively. 

\subsubsection{Ion-neutral damping}

The frequency of ion--neutral collisions $\nu_c$, which determines the rate of momentum transfer from ions to neutrals, is given by \citep{kulsrud71,drury96}
\begin{equation}
\nu_c=n_n \langle\sigma v_{th}\rangle=2\,n_n\,8.4\times10^{-9}\left(\frac{T}{10^4 K}\right)^{0.4}\,\rm s^{-1},
\end{equation}
where $\langle\sigma v_{th}\rangle$ is the fractional rate of change of the proton velocity $v_{th}$ averaged over the velocity distribution, $T$ is the temperature, and $n_n$ is the density of neutrals.

\begin{figure}
\includegraphics[scale=0.42]{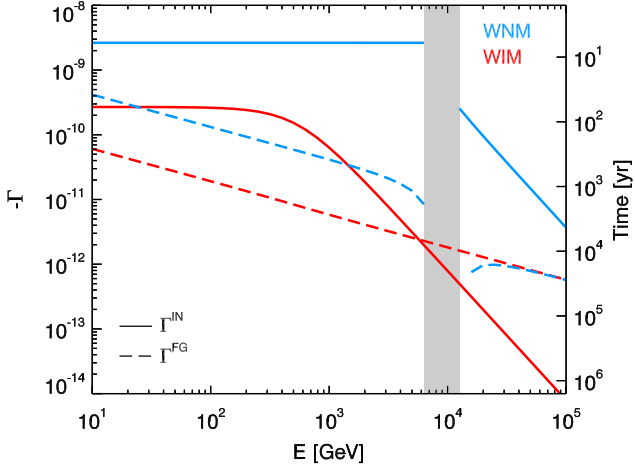}
\caption{Ion-neutral damping rate ($\Gamma^{IN}_{\rm d}$, solid curves) and Farmer \& Goldreich damping rate ($\Gamma^{FG}_{\rm d}$, dashed curves) as a function of the energy of resonating CRs. Different colours refer to two different phases of the interstellar medium: warm ionised medium (WIM, red), and warm neutral medium (WNM, blue). The shaded region refers to the WNM and indicates the range where Alfv\'en waves do not propagate.}
\label{fig:damp}
\end{figure}

The damping rate depends on how this frequency compares with the wave frequency $\omega_k$ that an \aw\ wave with wavenumber $k$ would have if there were no interactions between ions and neutrals: $\omega_k\equiv kB_0/\sqrt{4\pi m_pn_i}$ (valid in the case of slab modes), and on the ratio $\varepsilon$ between the ion density $n_i$ and the neutral density $n_n$: $\varepsilon=n_i/n_n$.
We estimate the ion-neutral damping rate by numerically solving equation A4 in \cite{zweibel82}. 
The results are shown in Fig.~\ref{fig:damp} (solid lines), as a function of the energy of the resonating particles, for both of the ISM phases considered.

Two asymptotic behaviors can be identified at small and large $k$, and they can be understood as follows:
\begin{itemize}
\item  $\omega_k\ll\nu_c$: at low wave frequencies ions and neutrals are well coupled  and the \aw\ velocity should be estimated using the total ISM density: $V_{\rm A}=kB_0/\sqrt{4\pi m_p n_{tot}}$. The damping rate is then $\Gamma^{IN}_{\rm d}\simeq-\frac{\omega_k^2}{2\nu_c\varepsilon^2}\propto E_k^{-2}$,where $E_k$ is the energy of particles resonating with waves of wavelength $\omega_k$.
\item $\omega_k\gg\nu_c$: the damping is more effective in this case, because ions and neutrals are not coupled and their collisions strongly damp the waves. Since neutrals do not participate to the coherent oscillations of the ions on the \aw\ waves, the \aw\ velocity is $V_{\rm A}=kB_0/\sqrt{4\pi m_p n_i}$. The damping rate is given by $\Gamma^{IN}_{\rm d}=-\frac{\nu_c}{2}\frac{\omega_k^2}{\omega_k^2+\nu_c^2\varepsilon^2}$, which for large $\omega_k$ is independent on the wave frequency: $\Gamma^{IN}_{\rm d}=-\nu_c/2$.
\end{itemize}

\begin{figure*}
\includegraphics[scale=0.48]{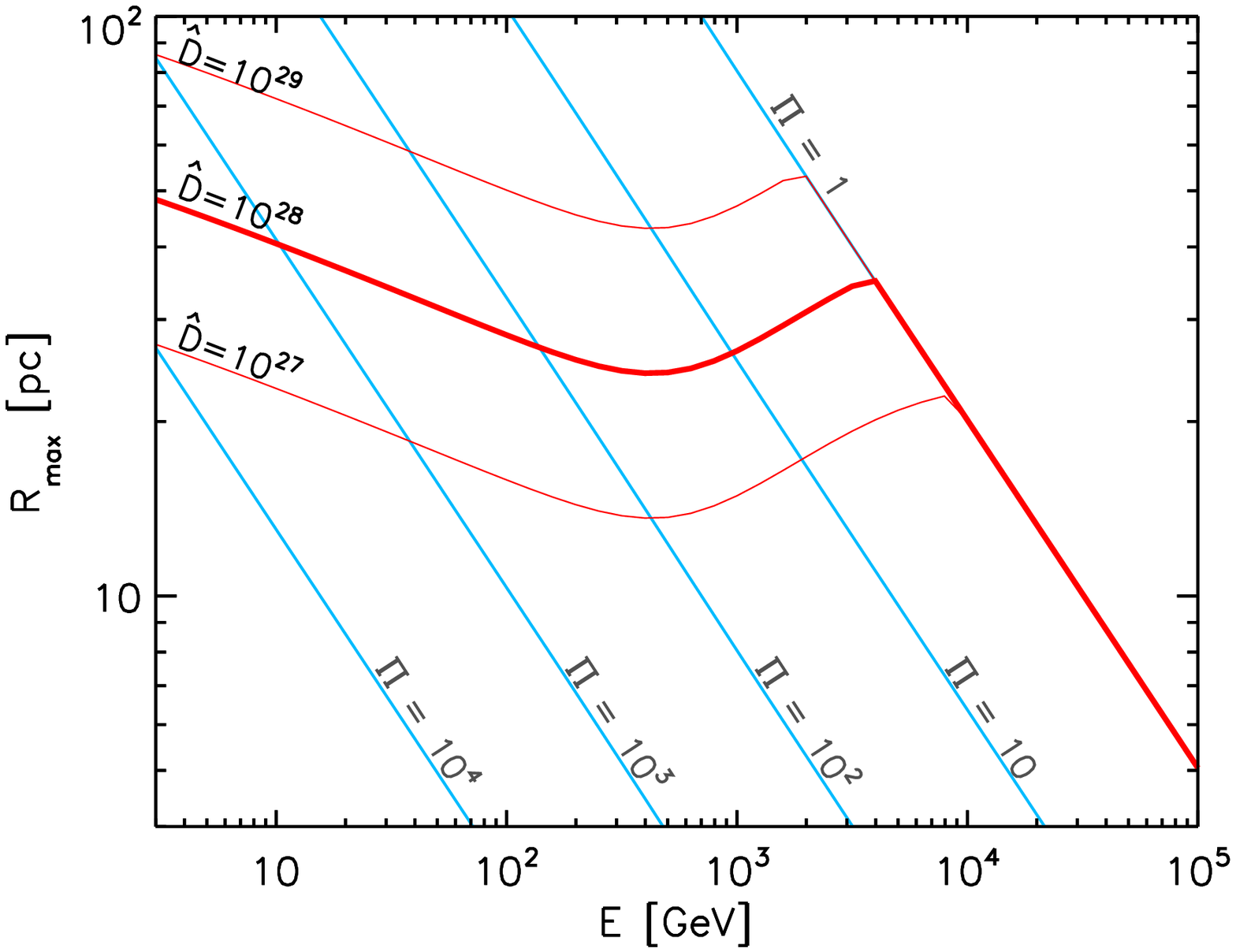}
\includegraphics[scale=0.48]{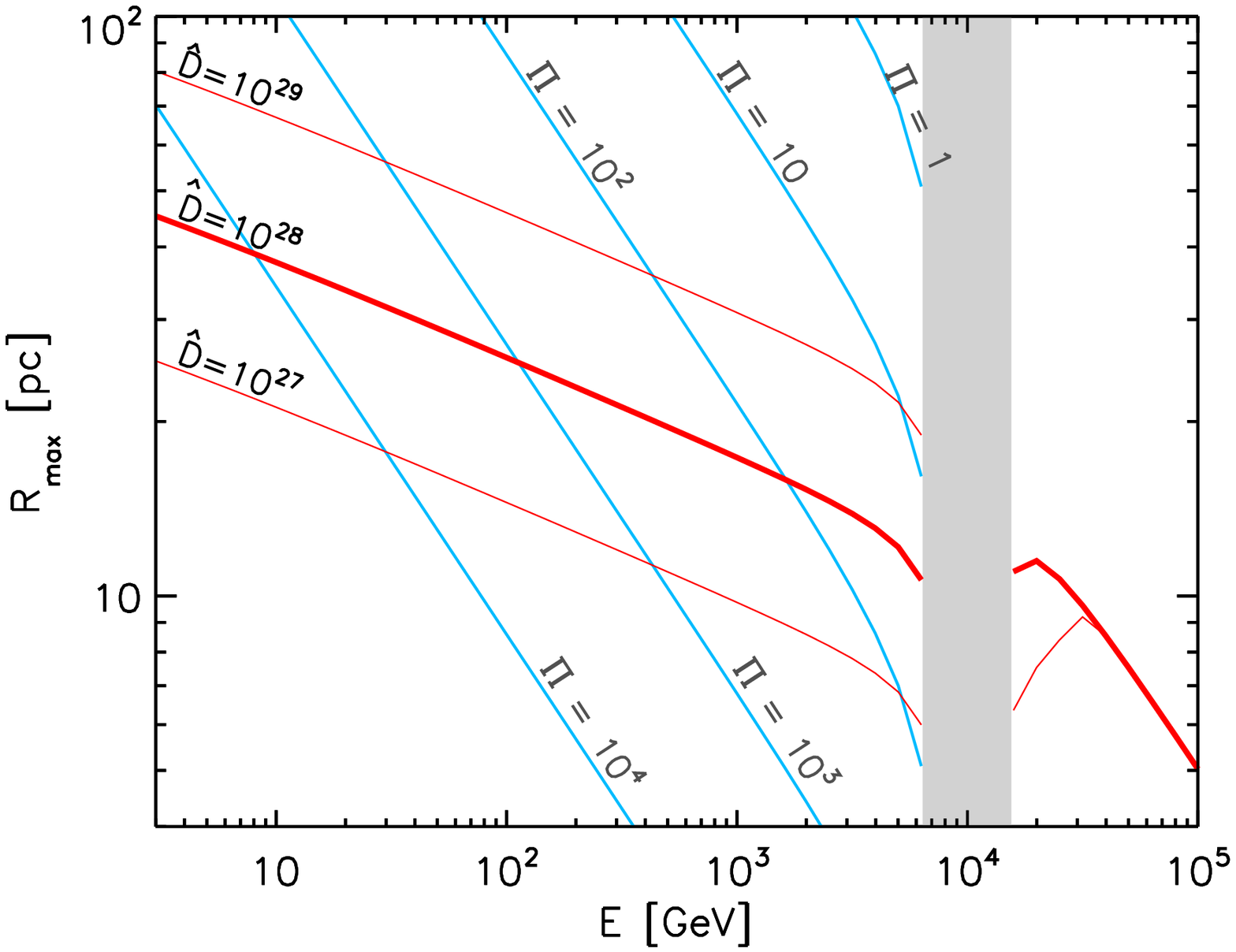}
\caption{Condition for the growth of Alfv\'en waves as given by the most stringent among equation \ref{eq:one} and \ref{eq:two}. Resonant Alfv\'en waves grow if protons of energy $E$ are released in the interstellar medium by an SNR of radius smaller than $R_{max}$. $R_{max}$ is represented by the red lines for three different values of the ambient (pre-existing) diffusion coefficient $\hat{D}_{28} =$ 0.1, 1, and 10 and for $W_{\rm CR,50} = 1$. It is given by the right-hand side parts of the inequalities in equation  \ref{eq:one} and \ref{eq:two} (see the text for more details). The left-hand (right-hand) panel refers to the WIM (WNM) phase of the interstellar medium. The values of the parameter $\Pi$ (equation~\ref{eq:pinum}) are also indicated with cyan lines.}
\label{fig:Rmax}
\end{figure*}

Note that if $\varepsilon<1/8$ (i.e. for a mostly neutral medium) there is a range of $k$ for which the waves do not propagate \citep{zweibel82}. This corresponds to a range of CR energies which is around $10^4\,$GeV for the parameters we chose to describe the WNM, and is indicated by a shaded region in Fig.~\ref{fig:damp}.
In the mostly neutral, weakly ionized atomic phases of the ISM, wave dissipation due to collisions of ions with neutrals can play an important role and quickly damp the Alfv\'enic turbulence \citep{kulsrud69,kulsrud71}.
The time-scale over which the damping operates is shown on the $y$-axis on the right of Fig.~\ref{fig:damp}.

\subsubsection{Farmer-Goldreich damping}

Large-scale magnetohydrodynamic (MHD) turbulence can be injected by SNRs, stellar winds or galactic arms \citep{maclow04}.
Phenomenological models of Alfv\'enic MHD turbulence have found that the cascade from large to small scales is anisotropic; the energy cascades preferentially perpendicular to the magnetic field \citep{goldreich95}.
The cascade process is due to the interaction of counter-propagating Alfv\'en wave packets. However, this process is not restricted to waves that are part of the cascade but also to any Alfv\'en wave that can interact with cascading waves. Especially, this is true for the waves generated by CRs escaping an SNR.

As we consider only parallel propagating waves in resonance with CRs of Larmor radius $r_{\rm L}$ here, the damping rate is given by (\citealt{fg}, see their equation9):
\begin{equation}
\Gamma_{\rm d}^{FG}= \left({\epsilon \over r_{\rm L} V_{\rm A}}\right)^{1/2} \ .
\end{equation}
where $\epsilon = V_{\rm A}^3/L_{inj}$ represents the energy cascade rate per unit mass. $L_{inj}$ is the turbulence injection scale; fixed at 50 pc for both WIM and WNM phases (Yan \& Lazarian 2004). We obtain the expression:
\begin{equation}
\label{eq:farmer}
\Gamma_{\rm d}^{FG}= {V_{\rm A} \over (r_{\rm L} L_{inj})^{1/2}} \ .
\end{equation}
which is plotted in Fig.~\ref{fig:damp} as a dashed line for both the WNM and WIM of the ISM.

Fig.~\ref{fig:damp} shows that in the WIM the ion-neutral damping dominates up to particle energies of several TeV. At larger energies, the ion-neutral damping becomes less effective, and as a consequence, the Farmer and Goldreich damping becomes the most relevant mechanism. However, at such large energies the number of CRs is quite small and thus the effectiveness of streaming instability is reduced accordingly. On the other hand, in the WNM, the ion-neutral damping is always the dominant damping mechanism for all the energies relevant for this study. 

For completeness, we recall that ion-neutral friction also contributes to damp the background turbulence $W_0$, and thus can indirectly affect the estimate of the damping term in Equation \ref{eq:farmer}. A lower limit in the resonant energy of interacting particles should be considered, at an energy obtained after balancing  the cascade rate of background turbulence $V_{\rm A}/r_{\rm L}$ with the damping time $\Gamma_{\rm d}^{IN}$. This effect has been neglected here, since we treated the background turbulence essentially as a free parameter of the problem. For a detailed discussion of this issue, the reader is referred to \citet{xu}.

\subsection{Conditions for wave growth}

As seen in Section~\ref{sec:problem}, the condition for the growth of Alfv\'en waves is $\Pi > 1$ in the absence of wave damping, or $\Pi > \tau_{diff}/\tau_{damp}$, when damping is effective. It is convenient to rewrite these two conditions as constraints on the SNR radius at the time of particles escape. This gives:
\begin{equation}
\label{eq:one}
R_1 < 2~W_{\rm CR,50}^{1/2} ~n_{i,-1}^{-1/4} E_1^{-0.6}
\end{equation}
and
\begin{equation}
\label{eq:two}
R_1 < 6 ~W_{\rm CR,50}^{1/4} ~n_{i,-1}^{-1/8} ~\Gamma_{d,-2}^{-1/4} ~ \hat{D}_{28}^{1/4}~E_1^{-0.35},
\end{equation}
respectively, where the damping coefficient, which might well be a function of energy, has been normalised as $\Gamma_{\rm d} = 0.01 ~ \Gamma_{\rm d,-2}$~yr$^{-1}$.
Alfv\'en waves grow when the most stringent among Equation \ref{eq:one} and \ref{eq:two} is satisfied. 
In deriving Equation~\ref{eq:two}, we assumed that the CR diffusion coefficient in the pre-existing turbulence is equal to $D_0 = D_{\rm B}/I_0 = \hat{D} E_1^{0.5}$ with $\hat{D} = 10^{28} \hat{D}_{28}$~cm$^2$/s.
Note that its value is very uncertain. Estimates of the energy index for the average Galactic value vary over the range [0.3-0.7] (e.g. \citealt{strong_review}). Also, we are interested in the diffusion coefficient in the local environment of the SNR, which might be very different from the average Galactic one.

The maximum radius allowed by Equations \ref{eq:one} and \ref{eq:two} is plotted as red lines in Figure~\ref{fig:Rmax} for the WIM (left-hand panel) and WNM (right-hand panel) phases of the ISM. In each panel, the three red curves refer to the fiducial value of the pre-existing diffusion coefficient, $\hat{D}_{28} = 1$ (thick red curve) and to values of $\hat{D}_{28}$ ten times smaller and larger (two thin red curves). The total energy of CRs has been fixed to $W_{\rm CR,50} = 1$ and their differential energy spectrum is a power law of slope 2.2.
It is clear from this figure that the growth of waves is expected in a quite significant fraction of parameter space.
For particle energies smaller than $\approx 10^4$ GeV the condition expressed by Equation \ref{eq:two}, or $\Pi > \tau_{diff}/\tau_{damp}$, is more stringent, while for larger energies Equation \ref{eq:one}, or $\Pi > 1$, dominates. Note that the latter does not depend on the value of the pre-existing diffusion coefficient.
The value of the parameter $\Pi$ as a function of SNR radius and particle energy is also indicated with cyan curves.

\begin{figure}
\includegraphics[scale=0.45]{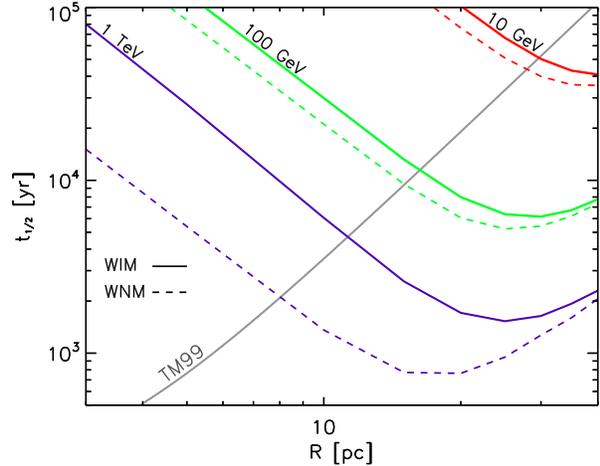}
\caption{Half-time of the CR cloud (see the text) as a function of its initial radius $R$. Red, green, and purple lines refer to particle energies of 10, 100, and 1000 GeV. Solid and dashed lines refer to the WIM and WNM phases of the interstellar medium. The total energy of CRs is set to $W_{\rm CR,50} = 1$. The black solid line represents the relationship between SNR radius and age according to \citet{truelove}.}
\label{fig:t1/2}
\end{figure}

Finally, following \cite{malkov}, we introduce the half-time of the CR cloud $t_{1/2}$. Let us assume that CRs of a given energy $E$ are initially uniformly distributed in a region (a cloud) of radius $R$, and let us follow the evolution in time of the CR cloud. The half-time of the CR cloud is defined as the time at which half of the CRs which were initially into the cloud have diffusively escaped.
The half-time is plotted in Figure \ref{fig:t1/2} as a function of the initial radius $R$ for particle energies of 10 GeV (red lines), 100 GeV (green lines), and 1 TeV (purple lines).
Solid and dashed lines refer to the WIM and WNM phases of the ISM, respectively.
Also in this case it has been assumed that the total energy of cosmic rays is set to $W_{\rm CR,50} = 1$ and that their differential energy spectrum is a power law of slope 2.2.

A characteristic trend is observed in Fig.~\ref{fig:t1/2}. At small radii, the half-time of the CR cloud decreases with the cloud's radius. 
In this regime, $\Pi$ is large, the waves growth quickly, and $t_{1/2}$ does not sensitively depend on the background turbulence $D_0$.
Larger radii correspond to smaller values of the parameter $\Pi$ (see Equation~\ref{eq:pinum}), resulting in a reduced efficiency in the amplification of Alfv\`en waves. This in turn translates in a less effective confinement of CRs, explaining why in this regime the half-time of the CR cloud decreases with the cloud's radius.
At even larger radii  ($> 25$ pc), an opposite trend is observed, namely, the half-time increases with radius. This is due to the fact that above a critical value of the initial cloud radius, the parameter $\Pi \propto R^{-2}$ drops below unity, and the test particle regime is recovered. In such a regime,
$t_{1/2}$ depends also on the value on the background turbulence throughout the well-known scaling $t_{1/2} \approx (R/D_0)^{1/2}$.

Here we propose to consider the half-time of the CR cloud as a rough estimate of the escape time of CRs from the region of initial size $R$. Though we are aware of the fact that the process of particle escape from SNRs is still poorly understood \citep[see][for seminal discussions on this issue]{ptuskinzirakashvili2005,bell2013}, we propose to extend this operational definition of escape time to SNRs also.
For this reason, we plot as a black line in Fig.~\ref{fig:t1/2} the relationship between age $t = 10^3 t_{\rm kyr}$~yr and radius $R_{\rm s}$ of an SNR expanding in a homogeneous medium, which reads \citep{truelove,ptuskinzirakashvili2005}:
\begin{equation}
\label{eq:Rs}
R_{\rm s} = 5.0 \left( \frac{E_{\rm SNR,51}}{n} \right)^{1/5} \left[ 1 - \frac{0.09 M_{ej,\odot}^{5/6}}{E_{\rm SNR,51}^{1/2} n^{1/3} t_{\rm kyr}} \right]^{2/5} t_{\rm kyr}^{2/5} ~ {\rm pc}
\end{equation}
where $E_{\rm SNR,51}$ is the supernova explosion energy in units of $10^{51}$~erg, $n$ the total density of the ambient ISM in cm$^{-3}$ and $M_{ej,\odot}$ the mass of the supernova ejecta in solar masses. The equation above is valid for times longer than $\approx 0.3 ~E_{\rm SNR,51}^{-1/2} M_{ej,\odot}^{5/6} n^{-1/3}$ kyr, while for earlier times an appropriate expression for the free expansion phase of the SNR evolution must be used \citep{chevalier1982}. Also, the validity of equation~\ref{eq:Rs} is limited to times shorter than $\approx 3.6 \times 10^4 E_{\rm SNR,51}^{3/14}/n^{4/7}$ yr, which marks the formation of a thin and dense radiative shell \citep{cioffi}.
Equation~\ref{eq:Rs} is plotted in Fig.~\ref{fig:t1/2} as a black line labelled TM99 for $E_{\rm SNR,51} = M_{ej,\odot} = 1$ and $n = 0.35$.
Within this framework, we can reproduce the qualitative result that higher energy particles escape SNRs earlier than lower energy ones \cite[see e.g.][and references therein]{gabiciescape}.
In particular, we find that for a supernova exploding in the WIM, CRs of energy 1 TeV, 100 GeV, and 10 GeV leave the SNR when its radius is $\approx$ 11, 16, and 30 pc, respectively. This corresponds to SNR ages of $\approx$ 4.7, 12, and 51 kyr. 
For the WNM and for the same particle energies the escape radii are $\approx$ 8, 15, and 28 pc, corresponding to SNR ages of $\approx$ 2.1, 9.5, and 44 kyr, respectively.
The estimated escape times and radii are not affected by the choice of $D_0$, since they fall into the regime where the cloud half-time $t_{1/2}$ is dominated by the self-excited turbulence, and not by the background one (see ~\ref{fig:t1/2}). However, for decreasing values of $D_0$ (i.e. large background turbulence), the transition to the TP regime (where $t_{1/2}\propto (R/D_0)^{1/2}$) occurs at smaller radii. An effect of $D_0$ on the estimated escape radii would however require very small values of $D_0$, from 10 to 100 times smaller than the one considered here (depending on the considered CR energy).

\section{Time evolution of a cloud of cosmic rays released by a supernova remnant}\label{sec:Tim}

\begin{figure}
{
\includegraphics[scale=0.43]{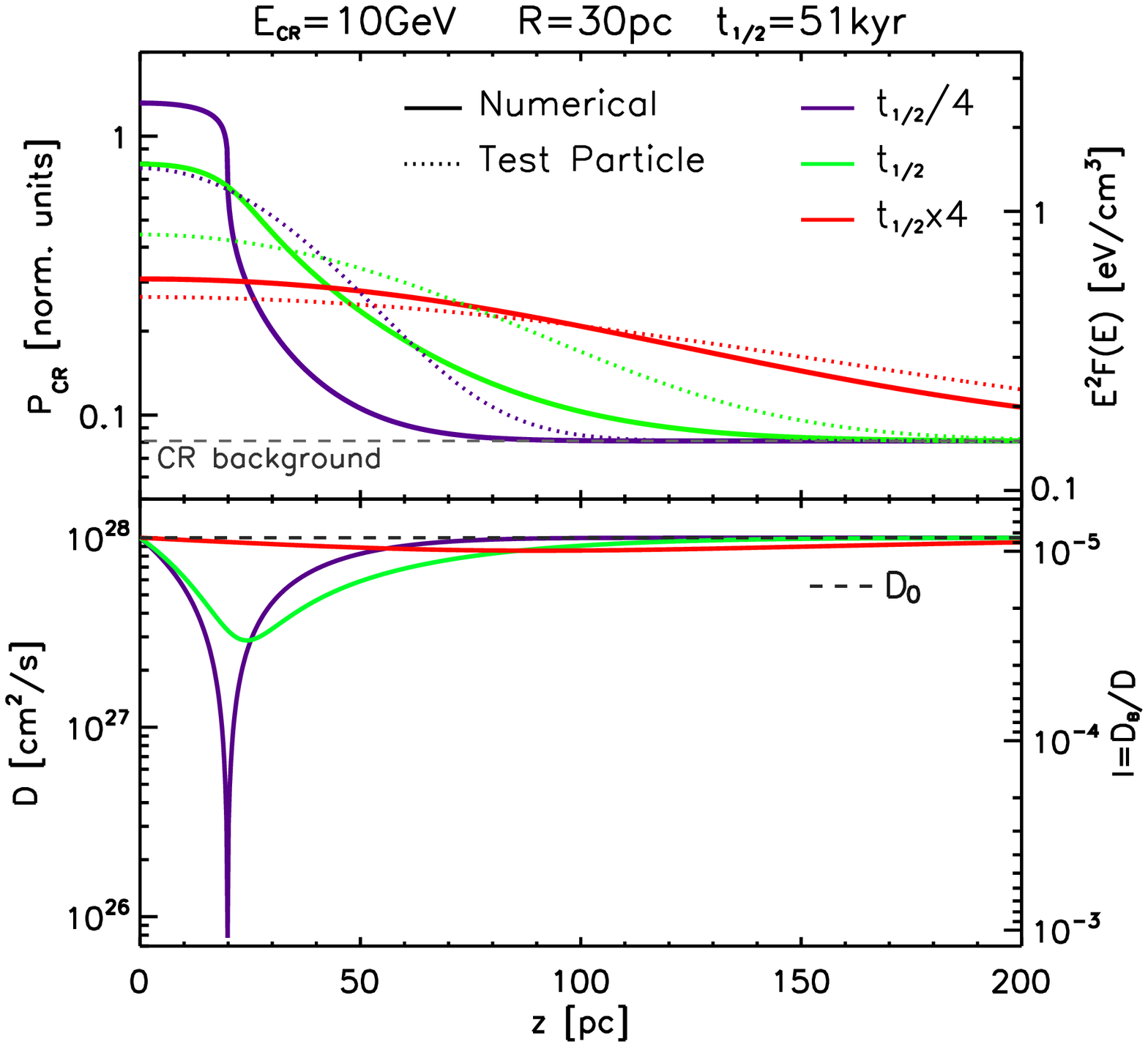}
\includegraphics[scale=0.43]{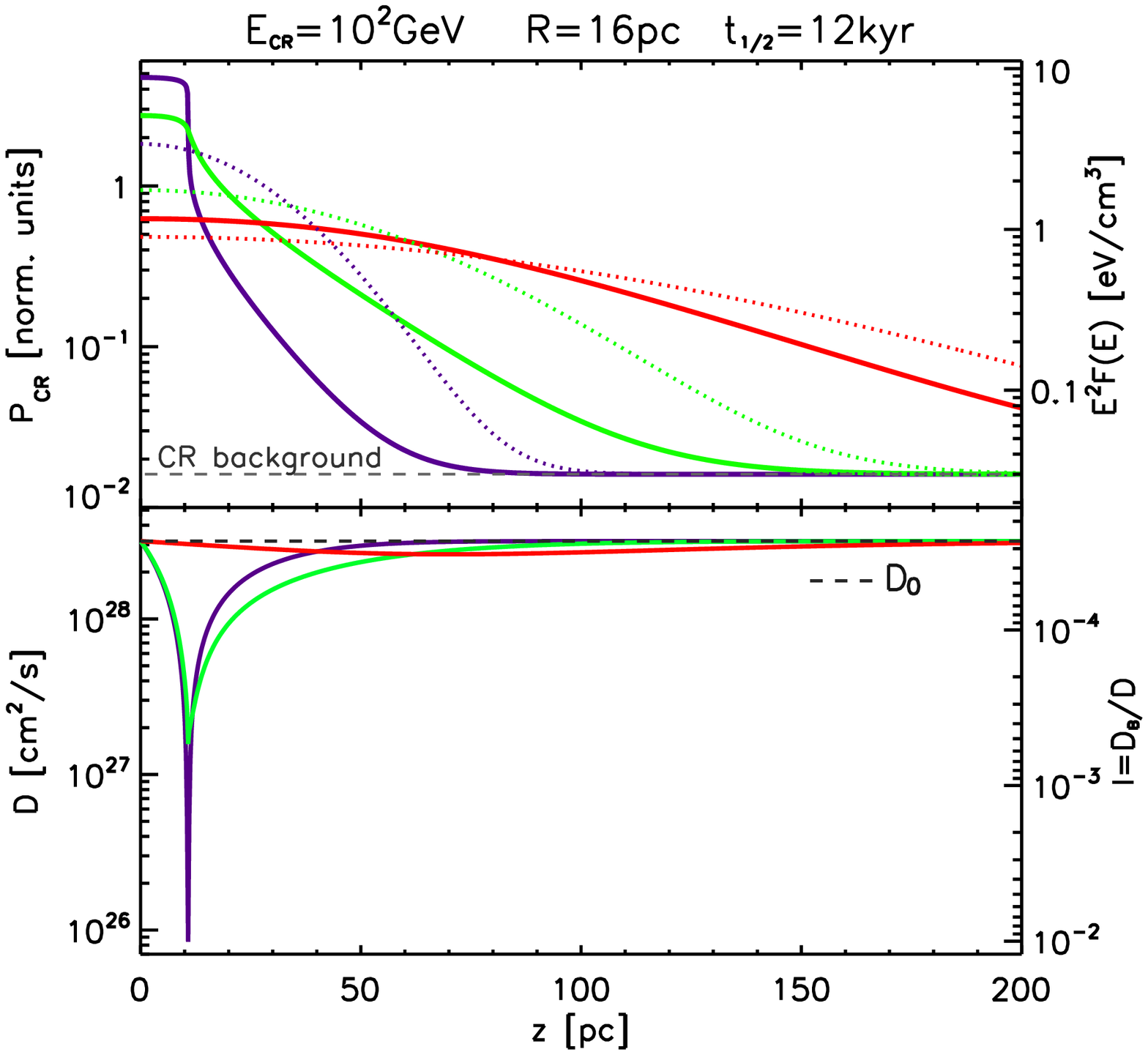}
\includegraphics[scale=0.43]{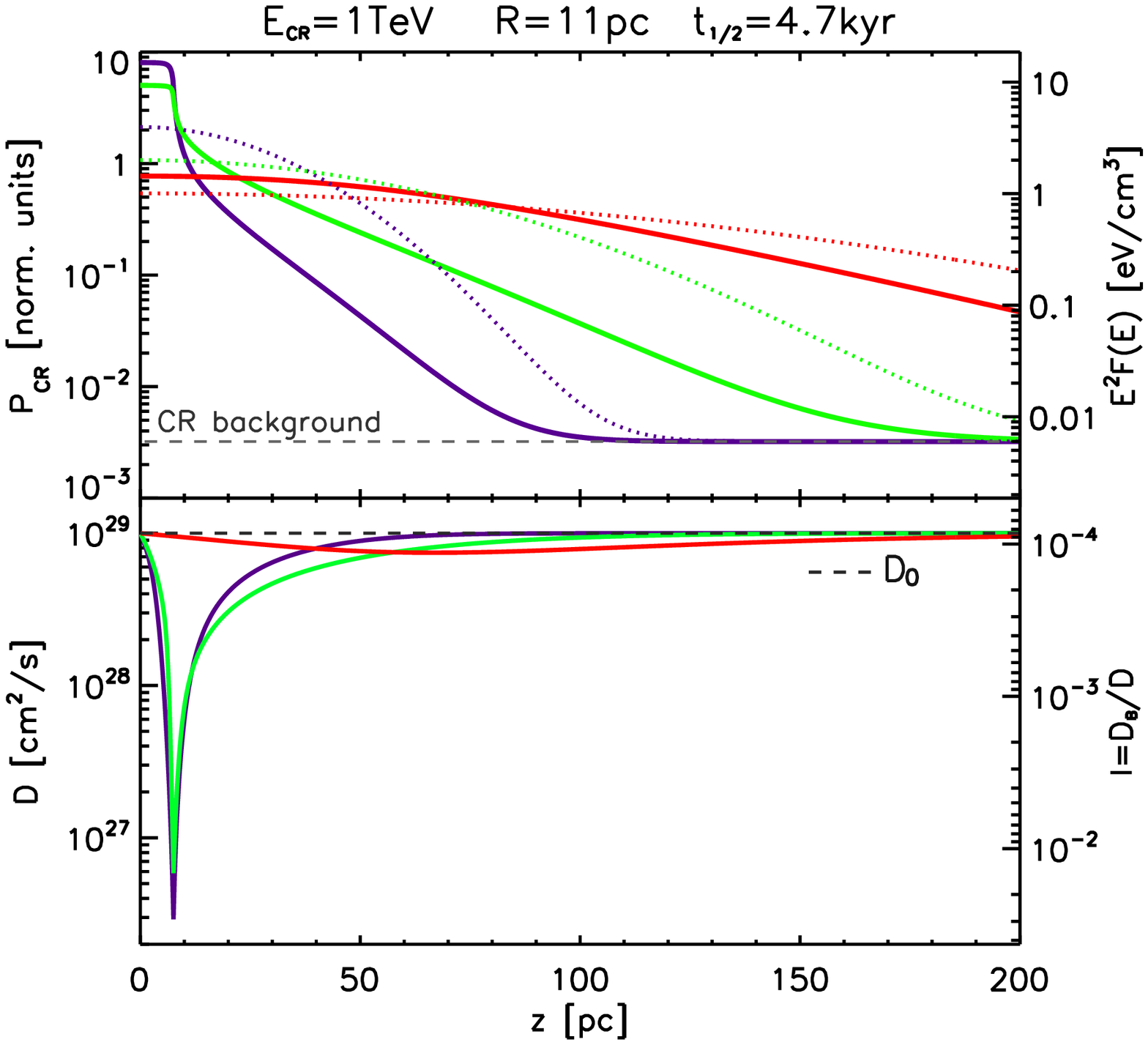}
}
\vskip -0.9 truecm
\caption{Time evolution of a CR cloud of initial radius $R$ in the WIM phase of the interstellar medium. CR energies of 10 GeV, 100 GeV, and 1 TeV are considered (top to bottom). The top (bottom) section of each panel shows the CR partial pressure (diffusion coefficient). See the text for more details.}
\label{fig:clouds}
\end{figure}

Fig.~\ref{fig:clouds} shows the evolution in time of a CR cloud. We assume that the CRs are released in the ISM by an SNR once the streaming instability is not effective enough to confine particles at the shock. According to the heuristic assumption described in the previous section, the intersections of the curves in Fig.~\ref{fig:t1/2} give an estimate of the escape time of CRs of a given energy (the half-life of the cloud $t_{1/2}$), and the corresponding radius of the SNR. In Fig.~\ref{fig:clouds} we considered the WIM phase of the ISM and the three panels refer to particles of energy of 10 GeV, 100 GeV, and 1 TeV (top to bottom). The half-life of the cloud and the escape radius are also indicated on the top of each panel. 
The upper section of each panel shows the CR pressure (in both normalised and physical units) as a function of the distance from the SNR centre, while the lower section shows the CR diffusion coefficient $D$, also in terms of ratio $I=D_{B}/D$ (right $y$-axis).
In all cases, $I\ll1$, and the assumption of quasi-linear theory is justified.
Purple, green, and red solid lines show the solution of Equations~\ref{eq:CRs} and \ref{eq:waves} at times equal to $t_{1/2}/4$, $t_{1/2}$, and $4 ~ t_{1/2}$, respectively, while dotted lines refer to the test-particle solution of the problem (i.e. streaming instability is not taken into account). The dashed black lines represent the level of the CR background in the Galaxy and the average turbulence level in the ISM $(\delta B/B)_k^2 = I_0(k)$ (upper and lower section of each panel, respectively).

\begin{figure}
\includegraphics[scale=0.45]{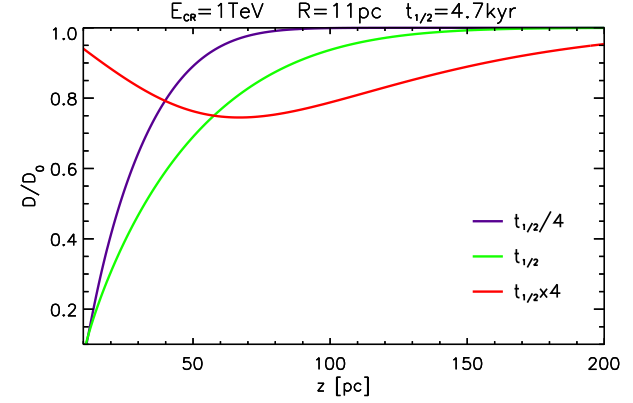}
\caption{Time evolution of the diffusion coefficient $D$ (normalised to the background galactic value $D_0$) for CR energies of 1 TeV. The initial radius and the half-time of the CR cloud are indicated on the top of the panel.}
\label{fig:diffusion}
\end{figure}

Several considerations are in order:
\begin{itemize}
\item{at early times, the solution of Equations~\ref{eq:CRs} and \ref{eq:waves} clearly differs from the test-particle solution, while for time-scales significantly longer than the half-time of the cloud $t_{1/2}$, the solution approaches the test particle one (see the red curves referring to a time equal to $4 t_{1/2}$). This implies that $t_{1/2}$ represents an order of magnitude estimate of the time interval during which waves can grow significantly above the background level in a region surrounding the initial CR cloud. This is an energy dependent effect, since $t_{1/2}$ is a decreasing function of energy. Thus, for CR energies of the order of 1 TeV or above, relevant for ground-based gamma-ray observations, the growth of waves operates for a quite short time interval (few thousands years or less);}
\item{large excesses of CRs above the galactic background can be maintained for times much longer than $t_{1/2}$. This is a well known result from the test-particle theory \citep[e.g.][]{atoyan,gabici09} which can be easily verified after comparing the values of the CR partial pressure (see scale on the right $y$-axis in Fig.~\ref{fig:clouds}) with the total energy density of CRs in the galactic disc, which is of the order of $\approx 1$~eV/cm$^3$;}
\item the CR diffusion coefficient is strongly suppressed with respect to its typical galactic value at the cloud border due to the strong gradient of CRs there, which translates in a fast growth rate of waves. The suppression of the diffusion coefficient remains significant in a region of tens of parsecs surrounding the SNR, as illustrated in Fig.~\ref{fig:diffusion}, where the diffusion coefficient of 1 TeV particles is plotted in units of the typical galactic value $D_0$. The ratio $D/D_0$ reaches a value equal to 0.5 at distances from the SNR centre equal to $\sim25\,$pc ($\sim35\,$pc) at times $t=t_{1/2}/4=1.2\,$kyr ($t= t_{1/2}=4.7\,$kyr), while at later time ($t=4\times t_{1/2}=19\,$kyr) the ratio is larger than 0.75 everywhere, reaching its minimum at a distance of $\sim65\,$pc;
\item A different assumption on the level of background turbulence would modify the solutions in the following way. Larger $D_0$ do not influence the region (typically at small distances/early times) where the turbulence is efficiently amplified, since for large amplifications, the evolution of $I$ becomes insensitive to the background level $I_0$. However, at larger distances, the self-excited turbulence decays quickly and the background turbulence dominates the diffusion. $D_0$ larger than those considered in Fig.~\ref{fig:clouds}, would in this case produce a faster transport, resulting in a higher CR pressure at larger distances. At late times the numerical solution approaches the TP solution, and its dependence on $D_0$ is well known.
Conversely, when a smaller $D_0$ is considered, the wave amplification is less relevant (because the background turbulence is already large) and the numerical solution tend to approach at all times the analytic solution. This can be seen from Fig.~\ref{fig:Rmax}: for small values of $\hat{D}$ the maximum radius above which the growth of Alfv\'en waves is not relevant is smaller.

\end{itemize}

\begin{figure}
\includegraphics[scale=0.49]{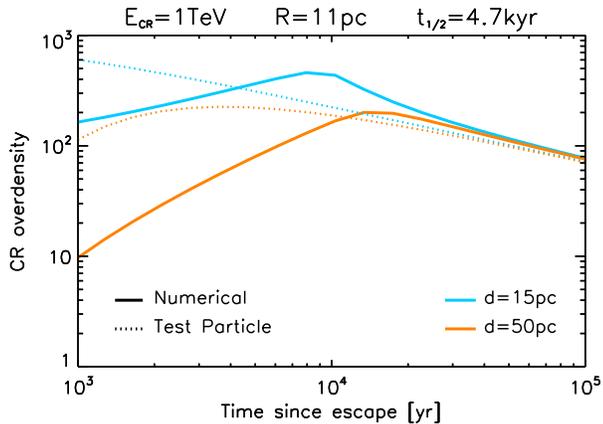}
\caption{Time evolution of the CR overdensity (defined as the ration between the energy density of CRs released by the source and the energy density of the CR background) at 1 TeV. Two different distances from the centre of the source are considered. Solid lines refer to the numerical solution, while dotted lines refer to the test particle approximation.}
\label{fig:overdensity}
\end{figure}

It is worth reminding that gamma-ray observations of molecular clouds located in the vicinity of SNRs have been used in order to constrain observationally the CR diffusion coefficient around SNRs \cite[e.g.][]{gabiciW28,nava13}. From an observational point of view, the main limitation of these studies, to date, is linked to the quite scarce number of SNR/molecular clouds associations currently detected in gamma-rays.
However, it is beyond any doubt that observations performed by next generation instruments (such as the Cherenkov Telescope Array) will increase significantly the statistics of detections and provide meaningful constraints to our model.
The gamma-ray emission from a molecular cloud is proportional to the mass of the molecular cloud and to the intensity of CRs at the cloud's location. 
In Fig.~\ref{fig:overdensity} the over-density of 1 TeV CRs above the galactic CR background is plotted as a function of time for a given distance from the SNR centre. 
This quantity is  proportional to the gamma-ray emission above $\sim 100$ GeV from a molecular cloud of a given mass located at a distance $d$ from the SNR.
The solid lines in Figure~\ref{fig:overdensity} refer to the results of our numerical study, while dotted lines represent the test-particle results. Cyan and orange lines refer to a distance from the centre of the SNR of $d =$ 15 and 50 pc, respectively.
It can be seen from the figure that at earlier times the intensity of CRs, and thus the corresponding gamma-ray emission, is suppressed with respect to the test particle case. This is because, in the presence of self-generated waves, CRs remain confined closer to the SNR than in the test-particle scenario, and reach the distance $d$ at later times only.
For the parameters considered here, the CR intensity becomes somewhat larger than that predicted by the test particle case at a time of the order of $\approx 10^4$ (which is approximately twice the half-time of the cloud), while at much larger times, our results coincide with the test-particle expectations, as expected.
This indicates that the interpretation of gamma-ray observations of molecular clouds located in the proximity of SNRs should be reconsidered in the light of our results. 
A detailed analysis of this issue goes beyond the scope of this study, and will be presented in a companion paper.

Finally, results very similar to those reported in Fig.~\ref{fig:clouds} are obtained for the WNM. This might seem at a first sight counter intuitive, given that the ion-neutral damping is much faster in the WNM than in the WIM (see Fig.~\ref{fig:damp}). 
However, this is balanced by the fact that the Alfv\'en speed, and thus the parameter $\Pi$ which regulates the initial growth of waves, is larger in the WNM (due to a smaller density of ions). We decided not to show the results for the WNM because they would look very similar to those reported in Fig.~\ref{fig:clouds}.

\section{Discussion and conclusions}
\label{sec:Con}

In this paper we presented a study of the propagation of CRs in the vicinity of SNRs. Following \citet{malkov}, we modelled this situation by considering a cloud of CRs subject to self-confinement.
In this scenario, Alfv\'enic turbulence is amplified by the streaming of CRs escaping the cloud. This in turn increases the confinement of CRs themselves around the cloud, making the problem highly nonlinear. The amplification of Alfv\'enic turbulence is balanced by various damping processes. In this paper, we focused mainly on the damping induced by the presence of neutral particles in the ambient medium, and for this reason, we limited out attention to SNRs expanding in the WIM and WNM phases of the ISM. A discussion of the case of SNRs expanding in the fully ionized regions of the galactic disc will be presented elsewhere (see also \citealt{plesser} and \citealt{dangelo}).

The main outcome of our study is the fact that, in the WIM and WNM, streaming instability affects the propagation of CRs after their escape from an SNR. The CR diffusion coefficient is significantly affected over a region of few tens of pc around the SNR, where its value can be suppressed by more than a factor of $\sim 2$ with respect to the average galactic value.
The suppression increases if smaller distances from the SNR are considered.
This implies that, when streaming instability is taken into account, the typical diffusion time of CRs up to a given distance is more than doubled with respect to the estimates from test-particle theory.
As pointed out in Section~\ref{sec:Tim}, this fact might have a great impact on the interpretation of the gamma-ray observations of molecular clouds located in the vicinity of SNRs, which have been often used to constrain the CR diffusion properties in the SNR environment \citep[see][and references therein]{gabicirev}. 

The half-time of the cloud, first introduced by \citet{malkov}, was found to be a crucial parameter of the problem. It represents the time it takes to have half of the CRs outside of the initial boundaries of the CR cloud, and it also gives a rough estimate of the time interval during which CR streaming instability amplifies the Alfv\'enic turbulence in the surrounding of the SNR. We estimated the half-time of the cloud for typical SNR parameters and we found it to be a decreasing function of particle energy. As a consequence, the propagation of $\approx 10$ GeV CRs is affected for several tens of kiloyears, while for $\approx 1$ TeV CRs, such time-scale reduces to few kiloyears.

Possible constraints on the scenario described in this paper might come from the detection of the gamma-ray emission generated by CR proton interactions with the ambient gas in the vicinity of SNRs. Such proton-proton interactions will also generate secondary electrons that will radiate through synchrotron and inverse Compton scattering from the radio to the hard X-ray domain, providing a multi-wavelength signature of particle escape from SNRs.


\section*{Acknowledgements}
LN was partially supported by a Marie Curie Intra-European Fellowship of the European Community's Seventh Framework Programme (PIEF-GA-2013-627715), 
by the ISF-CHE I-Core centre for Excellence for research in Astrophysics (1829/12), 
and by the China-NSF-Israel-ISF grant.

\label{lastpage}

\end{document}